\documentclass[epj,nopacs]{svjour}
\usepackage{graphicx}
\usepackage{color}
\usepackage{latexsym}
\usepackage[colorlinks,linkcolor=blue,citecolor=blue,urlcolor=black]{hyperref}

\newcommand{\be}{\begin{eqnarray}}
\newcommand{\ee}{\end{eqnarray}}

\begin{document}
\title{Testing Ghasemi-Nodehi-Bambi spacetime with continuum-fitting method}
\author{M. Ghasemi-Nodehi\inst{1} % etc
\thanks{\emph{email:} mghasemin@ipm.ir}%
}                     % Do not remove
\institute{School of Astronomy, Institute for Research in Fundamental Sciences (IPM), P. O. Box 19395-5531, Tehran, Iran}
\date{Received: date / Revised version: date}
\abstract{
The continuum-fitting method is the analysis of the thermal spectrum of the geometrically thin and optically thick accretion disk around stellar-mass black holes. A parametrization aiming to test the Kerr nature of astrophysical black holes was proposed in Ghasemi-Nodehi and Bambi in EPJC 76: 290, 2016. The metric contains 11 parameters in addition to the mass and spin parameters. One can recover the Kerr case by setting all parameters to one. 
 In this paper, I study the continuum-fitting method in Ghasemi-Nodehi-Bambi background. I show the impact of each of the  parameters on the spectra. I then employ $\chi^2$ studies and show that using the continuum-fitting method all parameters of Ghasemi-Nodehi-Bambi spacetime are degenerate. However, the parameter $b_9$ can be constrained in the case of a high spin value and the Ghasemi-Nodehi-Bambi black hole as reference.
 This degeneracy means that the spectra of the Kerr case cannot be distinguished from spectra produced in Ghasemi-Nodehi-Bmabi spacetime. This is a problem as regards measuring the spin of astrophysical black holes and constrain possible deviations from the Kerr case of General Relativity.
\PACS{
      {PACS-key}{discribing text of that key}   \and
      {PACS-key}{discribing text of that key}
     } % end of PACS codes
} %end of abstract
\maketitle
%

%%%%%%%%%%%%%%%%%%%%%%%%%%%%%%%

\section{Introduction}

General Relativity (GR) was born more than a century ago and is still used for describing the gravitational field and geometry of spacetime~\cite{einstein1916}. The theory has been extensively explored in the weak field regime, such as in solar system experiments and radio pulsars observations~\cite{will,will2,will3,will4}. There are many theoretical models that make the same prediction as general relativity in the weak field regime but they deviate in strong field regimes. The strong field regime can be either studied with gravitational waves (GWs) or with electromagnetic radiation. Dynamical systems have a signature on GW but an electromagnetic radiation method can be studied when spacetime is static. Effects such as the deviation from geodesics motion can be studied by electromagnetic radiation-related methods.

GR predicts astrophysical black holes (BHs) are Kerr BHs with spin and mass parameters \cite{kerr1,kerr2}. There are different scenarios that deviate from the Kerr case but there is degeneracy in measurement of their parameters and spin parameters of Kerr spacetime. 
One way to test the Kerr hypothesis is to parametrize the Kerr metric and try to constrain deviations from Kerr spacetime. There are several parametrization in the literature~\cite{p1,p2,p3,p4,p5,p6,p7,p8,p9,p10,p11,p12,p13,gbma,p14}. In this paper, I consider the Ghasemi-Nodehi-Bambi (GB) metric, proposed in~\cite{gbma}. The Kerr case would be recovered by setting all GB parameters to one. 
For an electromagnetics-based test of the strong gravity regime, one can use X-ray continuum spectra and iron line emission. One can also study BH shadow observations with an event horizon telescope at mm/sub-mm wavelength. In spite of large systematic errors of quasi-periodic oscillations (QPOs), one can use QPO frequencies to probe strong field regimes as well. Iron line reverberation mapping can be used as well.

Continuum spectra, iron lines and QPOs have already been observed by X-ray missions including RXTE, Chandra, Nustar, ASCA, Suzaku, and XMM-Newton.

The shadow~\cite{gbma}, X-ray reflection spectroscopy~\cite{gbiron}, QPO~\cite{qpoma} and iron line reverberation~\cite{qpoma} of the GB metric have already been studied. In the current paper, I study the continuum-fitting method to check possible constraints of the GB parametrization.

The continuum-fitting method is used to study the thermal spectrum of geometrically thin and optically thick accretion disks~\cite{confir}. One can only apply this method to stellar-mass BHs. 
This is because the spectrum of the supermassive BHs is in the optical/UV band where dust absorption limits the capability of accurate measurement. 
The temperature of the disk depends on the mass of the compact object. The spectrum of the thin disks around stellar-mass black holes is in the soft X-ray band. Currently, assuming the Kerr BH as astrophysical BH, the spin parameter, $a_*= J/M^2$, of the BH can be measured by this method. If one considers non-Kerr BHs, this technique can measure deviations from the Kerr solution~\cite{nk1,nk2}.

In this paper, I consider a GB background metric. I simulate thermal spectra of the thin accretion disk and show the impact of the deformation parameters on the spectra. By decreasing (increasing) the deformation parameter $b_7$ the spectra get softer (harder) and by decreasing (increasing) the parameters $b_4, b_9, b_{10}$ the spectra get harder (softer). The impact of the parameters $b_2, b_5, b_8, b_{11}$ is very close to Kerr one. 
The impact of the other parameters was studied before. References~\cite{nk1,con1,con3} provide the impact of all other parameters. By increasing (decreasing) the mass the spectra get harder (softer). The mass accretion rate changes the temperature of the disk. The distance of the source alters the normalization of the spectrum. For higher (lower) inclination angle, the area of the observed emitting region decreases (increases). If the spin parameter increases the spectra get harder.

Then, in order to compare results with the Kerr case, I use a minimum $\chi^2$ approach. I plot the related contour plots. The results show that using continuum-fitting method the GB parameters are degenerate and cannot be  distinguished  from the Kerr case. However, the parameter $b_9$ may be constrained in the case of a high spin value and the GB BH as a reference. 

The structure of the paper is as follows. In Sect.~\ref{con}, I provide details of the continuum-fitting method. Section~\ref{gb} is devoted to a description of the metric of the spacetime. Results and discussion are presented in Sect.~\ref{res}. Finally, a summary and conclusions are in Sect.~\ref{summary}.

%%%%%%%%%%%%%%%%%%%%%%%%%%%%%%%

\section{The continuum-fitting method \label{con}}

The thermal spectrum can be written in terms of the photon flux number density as measured by a distant observer~\cite{nk1,con1,con3,Li,boo}. The photon flux number density is given by
\be\label{eq-n2}
N_{E_{\rm obs}} =
\frac{1}{E_{\rm obs}} \int I_{\rm obs}(\nu) d \Omega_{\rm obs} 
\ee
where $I_{obs}$ is specific intensity of radiation, $E_{obs}$ is the photon energy, and  $\nu$ is the photon frequency measured by a distant observer. $d\Omega_{\rm obs} = dX dY / D^2$, $X$ and $Y$ are coordinate position of the photon on the sky as seen by a distant observer, $D$ is the distance of the source. 
In order to include all relativistic effects, one needs to compute the photon trajectory from the emission point in the disk to the image plane of the distant observer (detection point). We have
\be\label{eq-n22}
N_{E_{\rm obs}} &=&
\frac{1}{E_{\rm obs}} \int I_{\rm obs}(\nu) d \Omega_{\rm obs} = 
\frac{1}{E_{\rm obs}} \int w^3 I_{\rm e}(\nu_{\rm e}) d \Omega_{\rm obs}  \nonumber\\
\, 
\ee
where $I_e$ is the local specific intensity of the radiation emitted by disk and $w$ the redshift factor, which are
\be\label{eq-i-bb}
I_{\rm e}(\nu_{\rm e}) = \frac{2 h \nu^3_{\rm e}}{f_{\rm col}^4} 
\frac{\Upsilon}{\exp\left(\frac{h \nu_{\rm e}}{k_{\rm B} T_{\rm col}}\right) - 1} \, ,
\ee
\be\label{eq-red}
w = \frac{E_{\rm obs}}{E_{\rm e}} = \frac{\nu}{\nu_{\rm e}} = 
\frac{k_\alpha u^{\alpha}_{\rm obs}}{k_\beta u^{\beta}_{\rm e}}\, .
\ee
Regarding Eq.~\ref{eq-i-bb}, since the disk is in thermal equilibrium the emission is blackbody-like. One can define the effective temperature, $T_{\rm eff}$, $\mathcal{F}(r) = \sigma T^4_{\rm eff}$, where $\sigma$ is the Stefan-Boltzmann constant and  $\mathcal{F}(r)$ is the time averaged energy flux from Novikov-Thorne model~\cite{nov1,nov2}. 
The Novikov-Thorne model can be used to describe geometrically thin and optically thick accretion disks around the black hole. The assumption is that the disk is in the equatorial plane. The gas of the disk moves on a nearly geodesic circular orbit. $\mathcal{F}(r)$ can be written as
\be
\mathcal{F}(r) = \frac{\dot{M}}{4 \pi \sqrt{-G}} F(r) \,\,\, , \,\,\,\sqrt{-G} = \sqrt{\alpha^2 g_{rr} g_{\phi\phi}}
\ee
where $\dot{M}$ is the mass accretion rate, and $F(r)$ is
\be\label{eq-f}
F(r) = \frac{\partial_r \Omega}{(E - \Omega L_z)^2} \int_{r_{\rm in}}^r
(E - \Omega L_z)(\partial_\rho L_z) d\rho \, .
\ee
$E, L_z,$ and $\Omega$ are, respectively, the
conserved specific energy, the conserved $z$-component of the specific angular momentum, and the angular velocity for equatorial circular geodesics. $r_{\rm in}$ is assumed to be at the innermost stable circular orbit (ISCO). Actually the disk's temperature near the inner edge is high and non-thermal effects are not negligible. For this reason one can introduce a hardening factor or a color factor, $f_{\rm col}$ and the color temperature is $T_{\rm col} (r) = f_{\rm col} T_{\rm eff}.$ In Eq.~\ref{eq-i-bb}, $\nu_e$ is the photon frequency, $h$ is Planck's constant, $k_B$ is the Boltzmann constant and $\Upsilon$ is a function of the angle between the wavevector of a photon emitted by the disk and the normal of the disk surface. Here I consider $\Upsilon = 1$ for isotropic emission.\\
Regarding the $w$ in Eq.~\ref{eq-red}, $E_e = h \nu_e, \nu$ being a photon frequency measured by a distant observer, $k^{\alpha}$ is the 4-momentum of the photon, $u^{\alpha}_{\rm obs} = (-1,0,0,0)$ is the 4-velocity of the distant observer, and 
$u^{\alpha}_{\rm e} = (u^t_{\rm e},0,0, \Omega u^t_{\rm e})$ is the 4-velocity of 
the emitter. From  Liouville's theorem we have $I_{\rm e}(\nu_{\rm e})/\nu_{\rm e}^3 = I_{\rm obs} (\nu_{\rm obs})/\nu^3$. \\
Finally, the photon flux number density can be written as 
\be\label{eq-n222}
N_{E_{\rm obs}} &=&
A_1 \left(\frac{E_{\rm obs}}{\rm keV}\right)^2
\int \frac{1}{M^2} \frac{\Upsilon dXdY}{\exp\left[\frac{A_2}{w F^{1/4}} 
\left(\frac{E_{\rm obs}}{\rm keV}\right)\right] - 1} \, ,
\ee
where $A_1$ and $A_2$ are
\be
A_1 &=&  
\frac{2 \left({\rm keV}\right)^2}{f_{\rm col}^4} 
\left(\frac{G_{\rm N} M}{c^3 h D}\right)^2  
\nonumber\\ &=& 
\frac{0.07205}{f_{\rm col}^4} 
\left(\frac{M}{M_\odot}\right)^2 
\left(\frac{\rm kpc}{D}\right)^2 \, 
{\rm \gamma \, keV^{-1} \, cm^{-2} \, s^{-1}} \, , \nonumber\\
A_2 &=&  
\left(\frac{\rm keV}{k_{\rm B} f_{\rm col}}\right) 
\left(\frac{G_{\rm N} M}{c^3}\right)^{1/2}
\left(\frac{4 \pi \sigma}{\dot{M}}\right)^{1/4}  
\nonumber\\ &=& 
\frac{0.1331}{f_{\rm col}} 
\left(\frac{\rm 10^{18} \, g \, s^{-1}}{\dot{M}}\right)^{1/4}
\left(\frac{M}{M_\odot}\right)^{1/2} \, .
\ee
One finds
\be
u^t_{\rm e} = - \frac{1}{\sqrt{-g_{tt} - 2 g_{t\phi} \Omega - g_{\phi\phi} \Omega^2}} \, ,
\ee
from the normalization condition $g_{\mu\nu}u^{\mu}_{\rm e}u^{\nu}_{\rm e} = -1$ and thus
\be\label{eq-red-g}
w = \frac{\sqrt{-g_{tt} - 2 g_{t\phi} \Omega - g_{\phi\phi} \Omega^2}}{1 + 
\lambda \Omega} \, ,
\ee
where $\lambda = k_\phi/k_t$ is a constant of motion along the photon's path. All relativistic effects are encoded in the redshift factor $w$.

%%%%%%%%%%%%%%%%%%%%%%%%%%%%%%%
\section{Ghasemi-Nodehi-Bambi spacetime}\label{gb}

Reference~\cite{gbma} proposed a new parametrization to the Kerr metric. The purpose of this parametrization is to constrain possible deviations from a Kerr solution of GR. The Kerr case would be recovered when all deformation parameters are equal to one. While in other metrics the deformation parameters are additive and reduce to the Kerr case for vanishing deformation parameters. In this parametrization, 11 new parameters are introduced in front of any mass and/or spin term. Any deviation from one deforms the spacetime more/less than that of prediction of GR. The metric is as follows:
\be\label{eq-m}
ds^2 &=& - \left( 1 - \frac{2 b_1 M r}{r^2 + b_2 a^2 \cos^2\theta} \right) dt^2 
\nonumber\\ &&
- \frac{4 b_3 M a r \sin^2\theta}{r^2 + b_4 a^2 \cos^2\theta} dt d\phi 
+ \frac{r^2 + b_5 a^2 \cos^2\theta}{r^2 - 2 b_6 M r + b_7 a^2} dr^2 
\nonumber\\ &&
+ \left( r^2 + b_8 a^2 \cos^2\theta \right) d\theta^2 
\nonumber\\ &&
+ \left( r^2 + b_9 a^2 + \frac{2 b_{10} M a^2 r 
\sin^2\theta}{r^2 + b_{11} a^2 \cos^2\theta} \right) \sin^2\theta d\phi^2 \, .
\ee
Here we set $b_1 = b_3 = b_6 = 1$. $b_1$ is equal to one because it is the coefficient of mass and $b_3 = 1$ in the same way;  $b_3 a$ is the asymptotic specific angular momentum. $b_6$ is close to one from a solar system experiment. I do not consider these three parameters in my continuum-fitting method calculations.

%%%%%%%%%%%%%%%%%%%%%%%%%%%%%%%%

\section{Results and discussion}\label{res}

\begin{figure*}
\vspace{0.4cm}
\begin{center}
\includegraphics[type=pdf,ext=.pdf,read=.pdf,width=12.0cm]{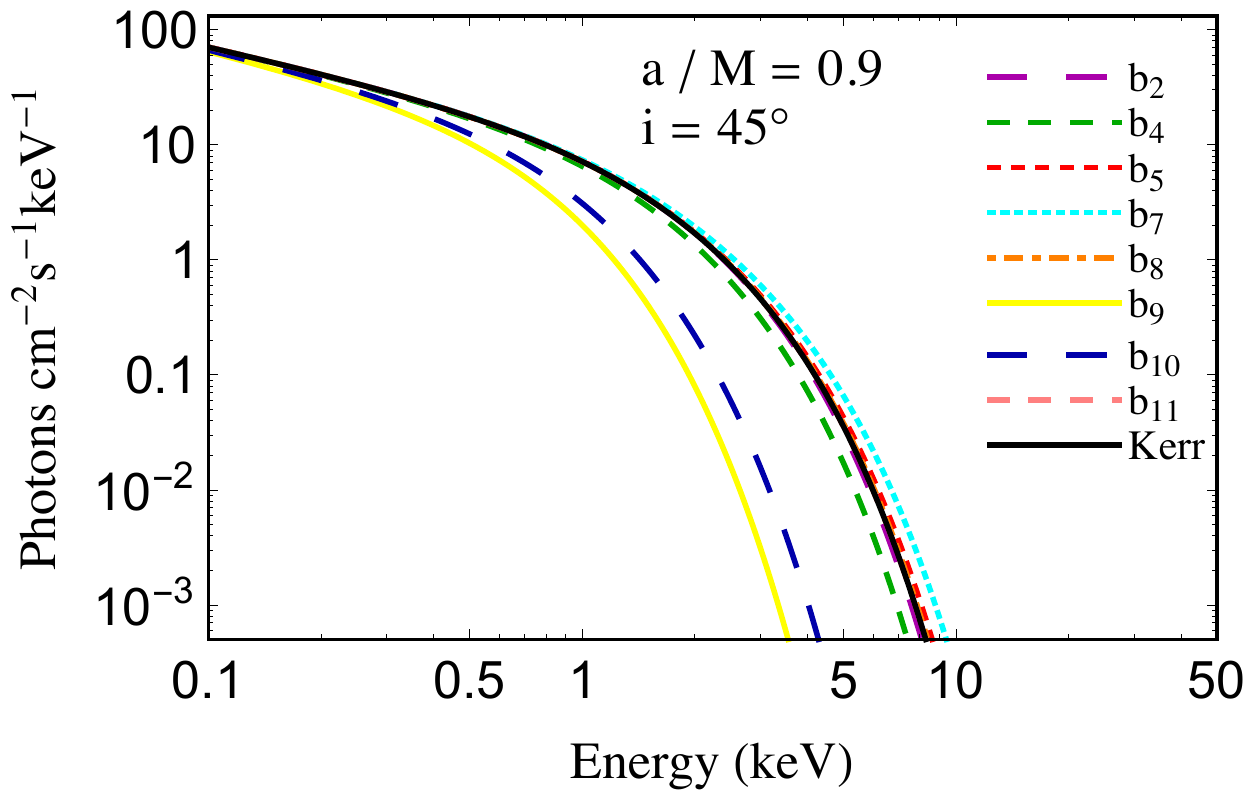}
\end{center}
\vspace{-0.3cm}
\caption{Thermal spectra of accretion disks for different values of the $b_i$ parameters. The values of the model parameters are $M = 10 M_{\odot}$,  $\dot M = 2 \times 10^{18}$ g/s, $D = 10$ kpc, $i = 45^{\circ}$, $a_* = 0.9,$ and one of $b_i = 15$ in each case. In these simulations I have assumed $f_{col} = \Upsilon = 1$. See text for more details.\label{spectra}}
\end{figure*}

\begin{figure*}
\vspace{0.4cm}
\begin{center}
\includegraphics[type=pdf,ext=.pdf,read=.pdf,width=8.0cm]{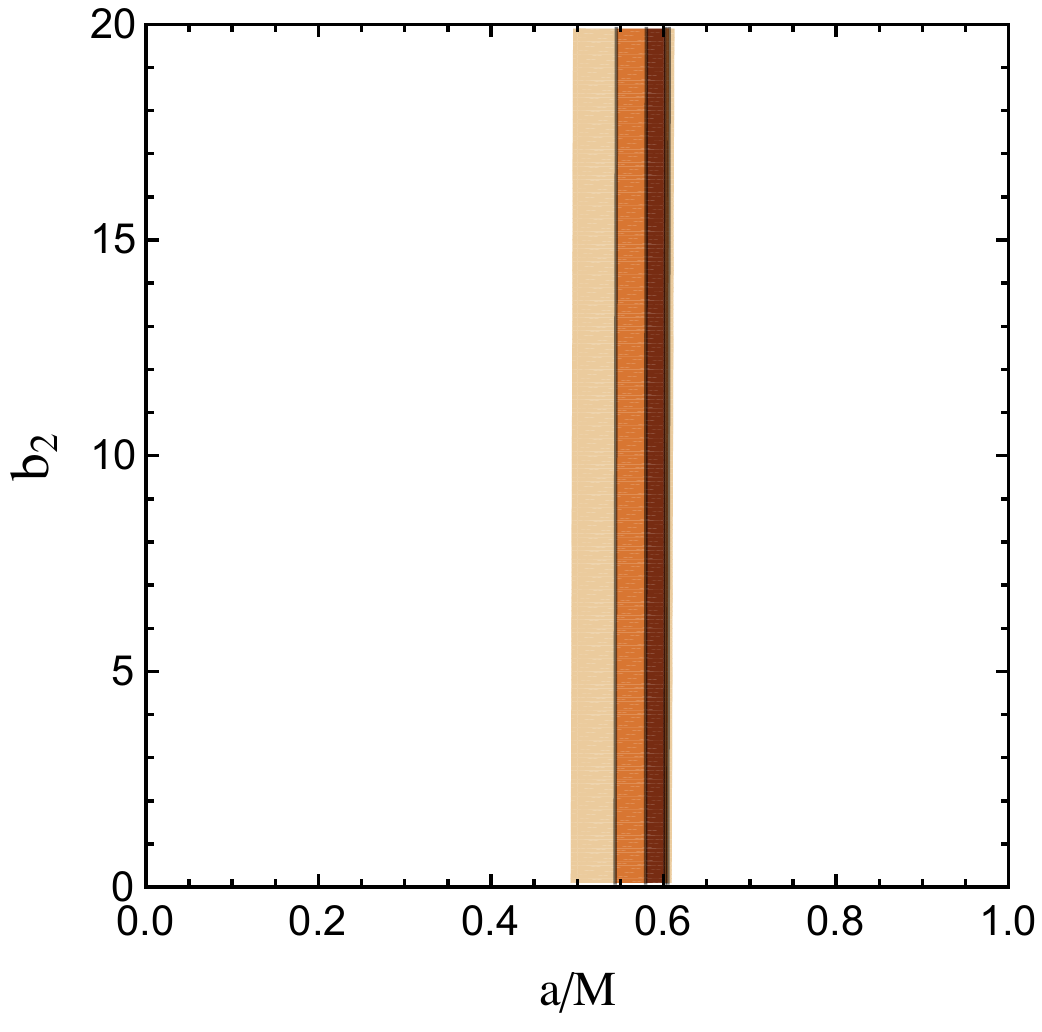}
\hspace{0.8cm}
\includegraphics[type=pdf,ext=.pdf,read=.pdf,width=8.0cm]{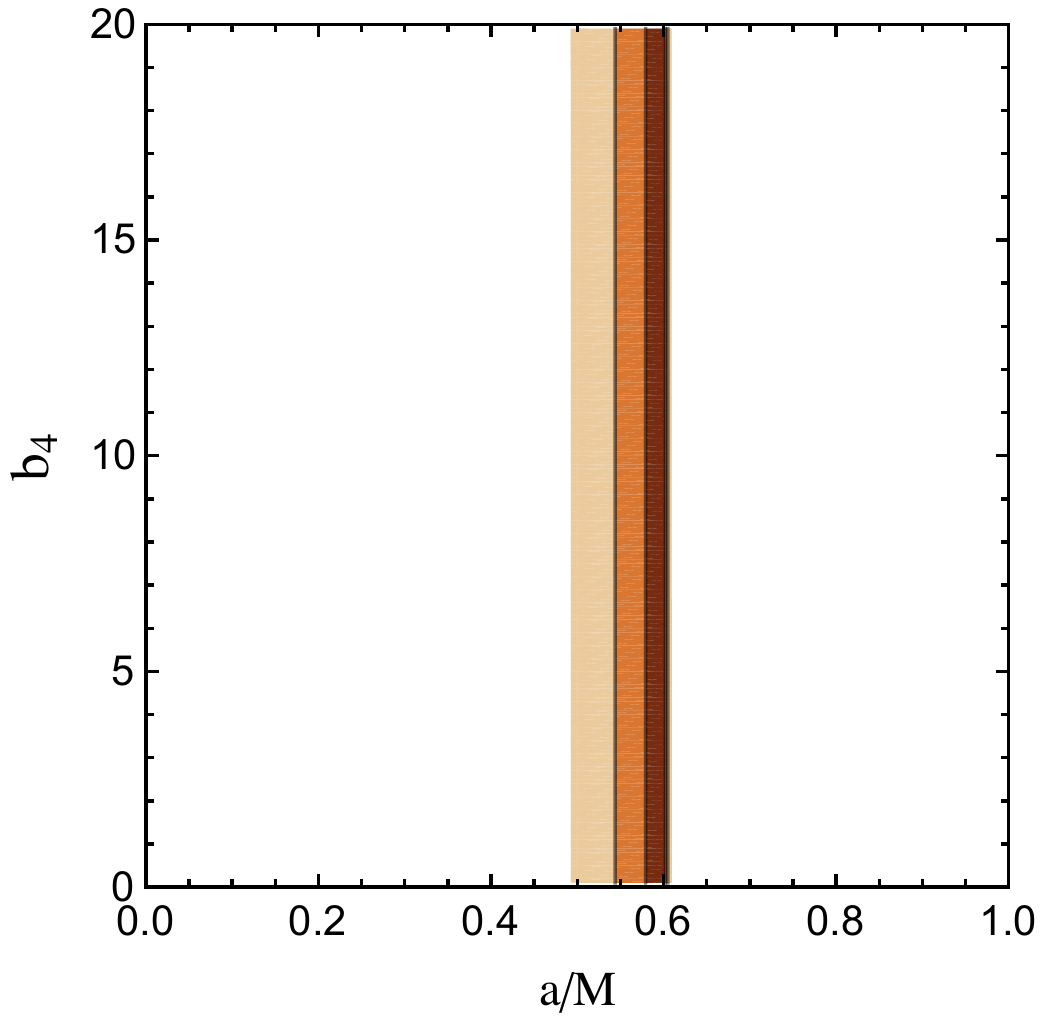}\\
\vspace{0.8cm}
\includegraphics[type=pdf,ext=.pdf,read=.pdf,width=8.0cm]{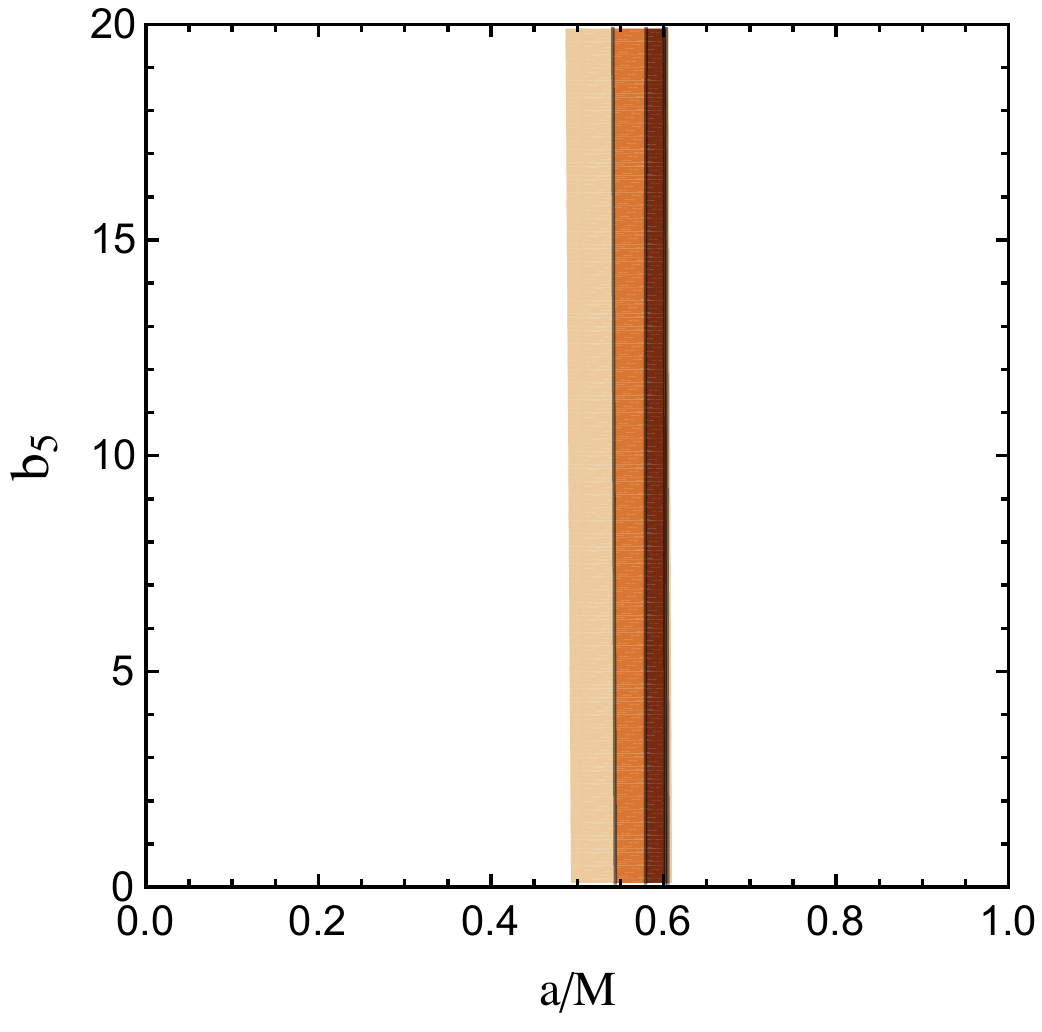}
\hspace{0.8cm}
\includegraphics[type=pdf,ext=.pdf,read=.pdf,width=8.0cm]{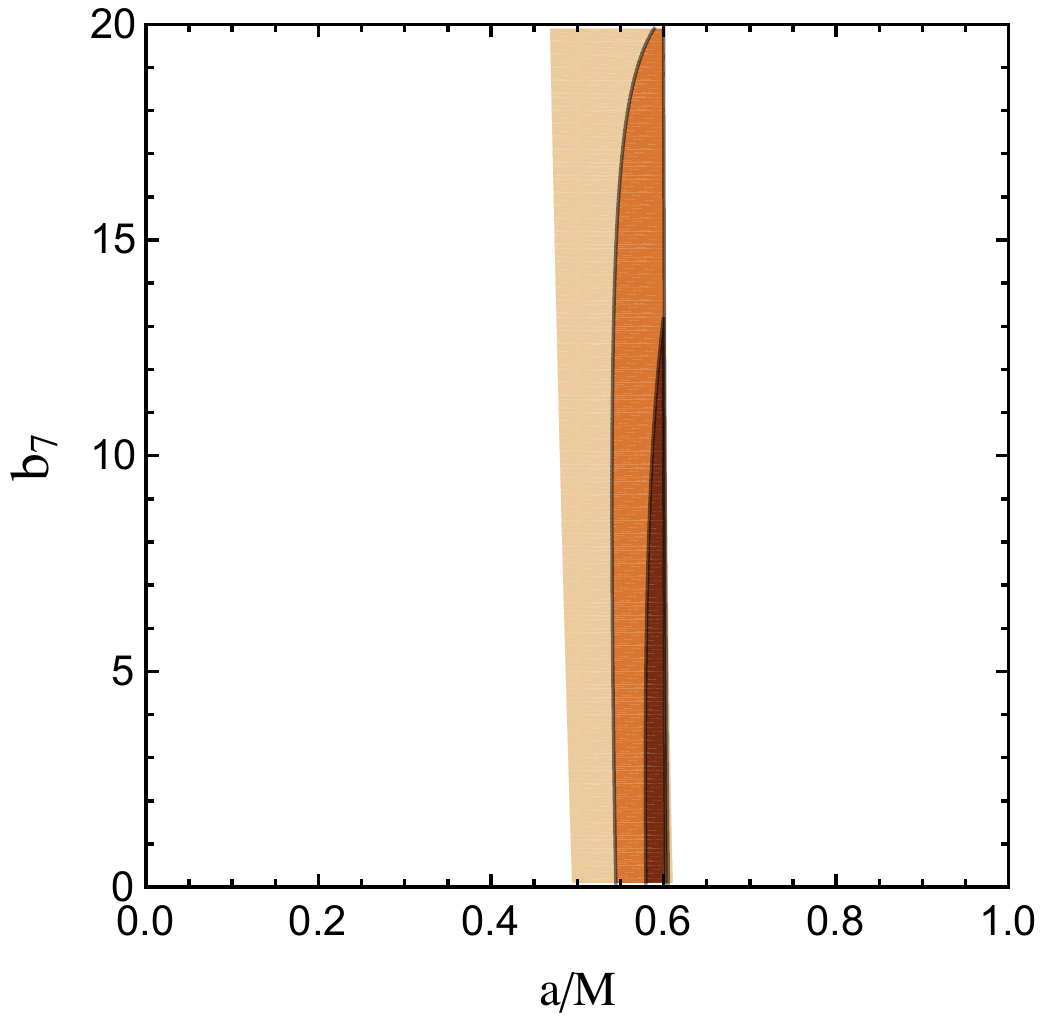}
\end{center}
\vspace{-0.3cm}
\caption{Contour map of thermal spectra of accretion disks. Contour map of parameter $b_2$ (top left panel), $b_4$ (top right panel), $b_5$ (bottom left panel), $b_7$ (bottom right panel). The reference model is a Kerr BH  with spin 0.6. The other parameter values are $M = 10 M_{\odot}$,  $\dot M = 2 \times 10^{18}$ g/s, $D = 10$ kpc, and $i = 45^{\circ}$. See the text for more details.  \label{con1}}
\end{figure*}

\begin{figure*}
\vspace{0.4cm}
\begin{center}
\includegraphics[type=pdf,ext=.pdf,read=.pdf,width=8.0cm]{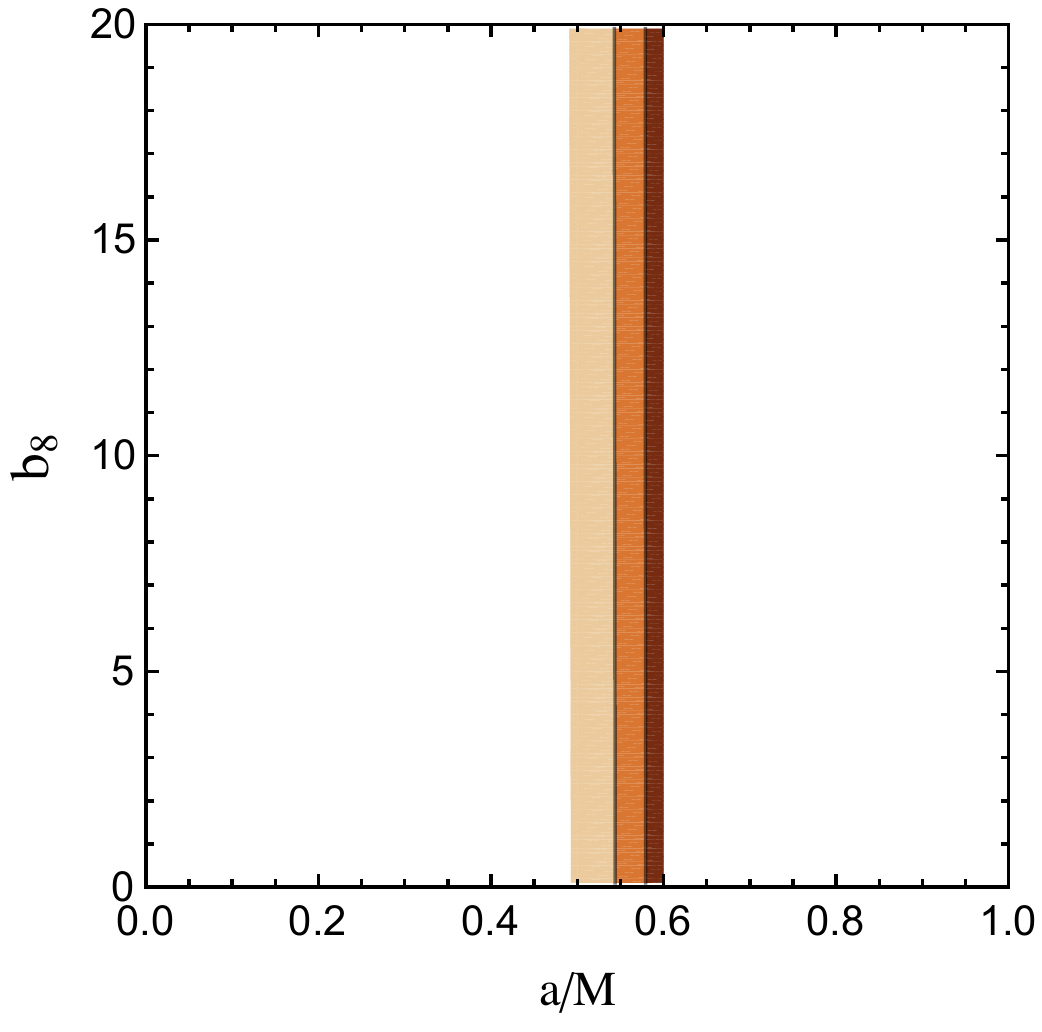}
\hspace{0.8cm}
\includegraphics[type=pdf,ext=.pdf,read=.pdf,width=8.0cm]{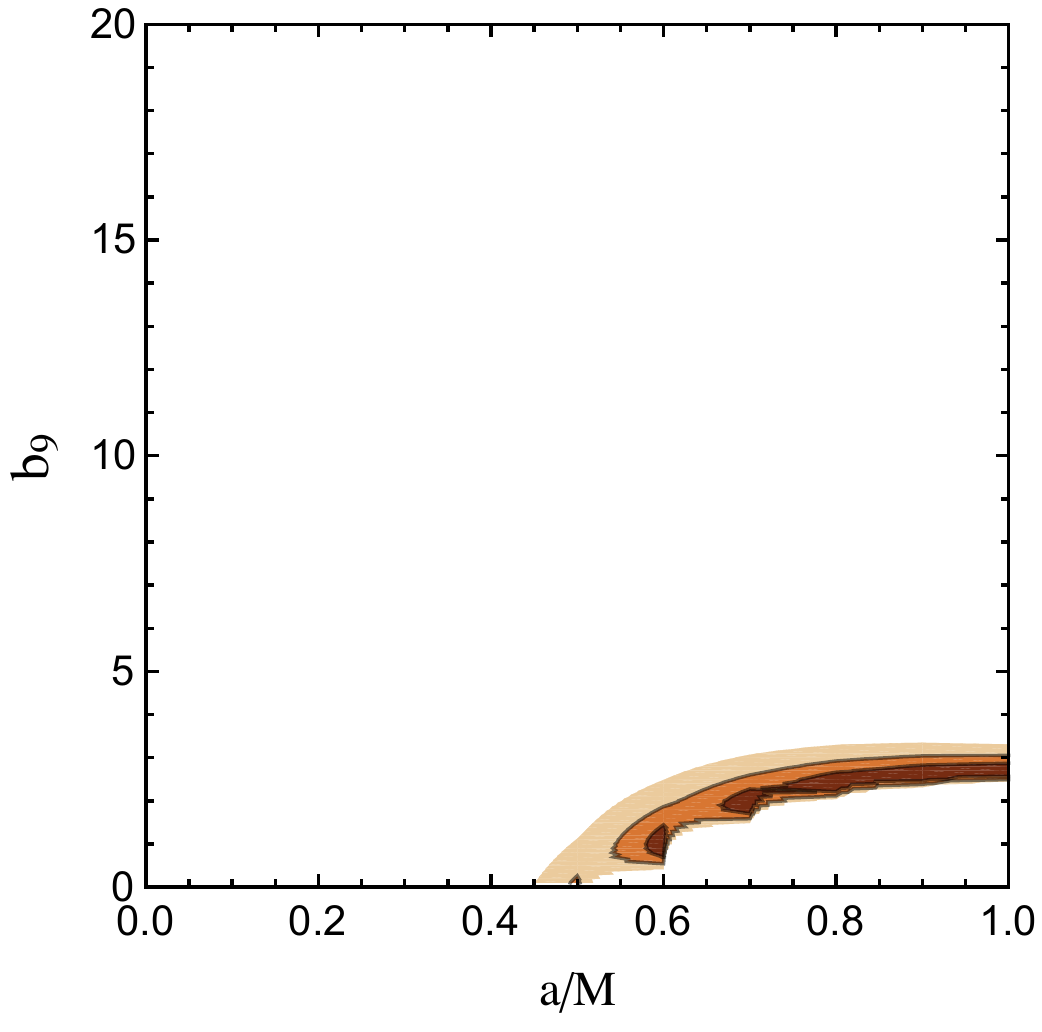}\\
\vspace{0.8cm}
\includegraphics[type=pdf,ext=.pdf,read=.pdf,width=8.0cm]{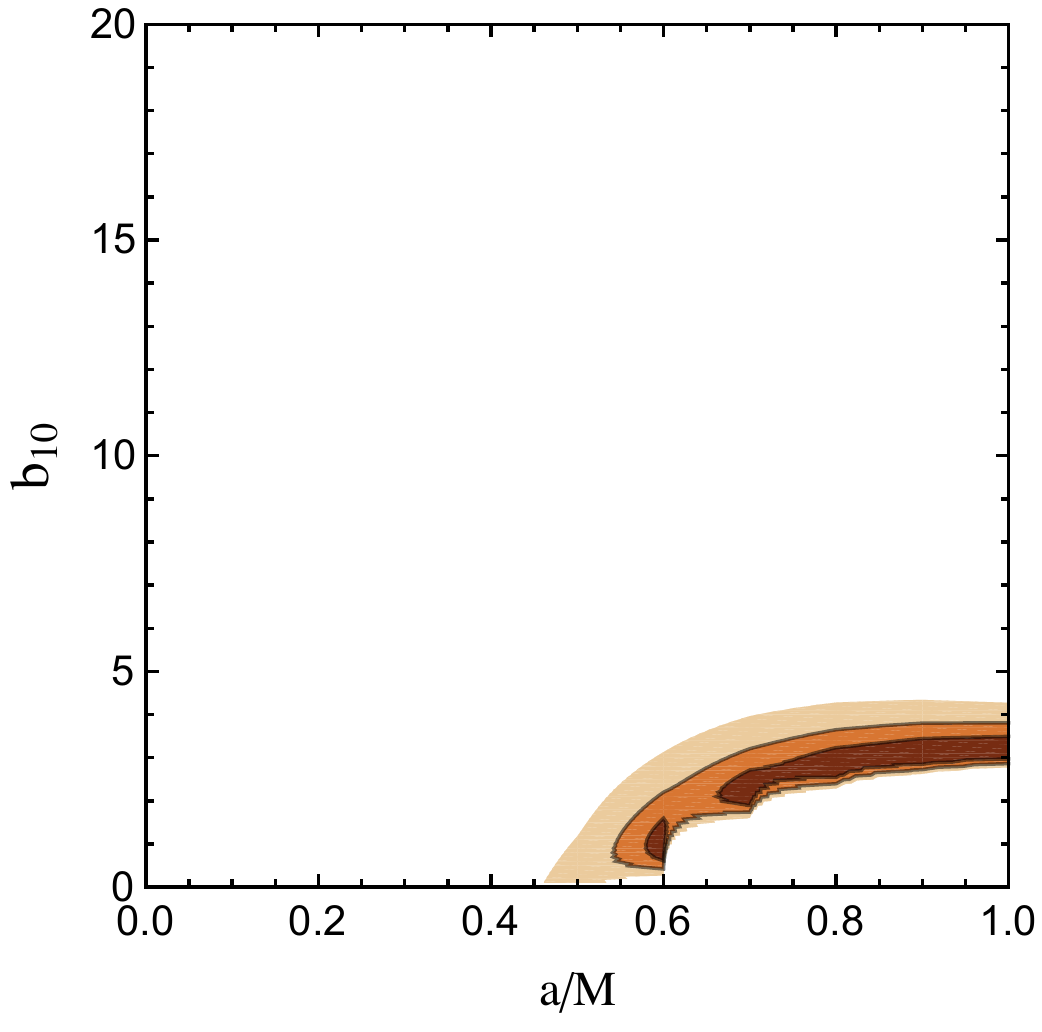}
\hspace{0.8cm}
\includegraphics[type=pdf,ext=.pdf,read=.pdf,width=8.0cm]{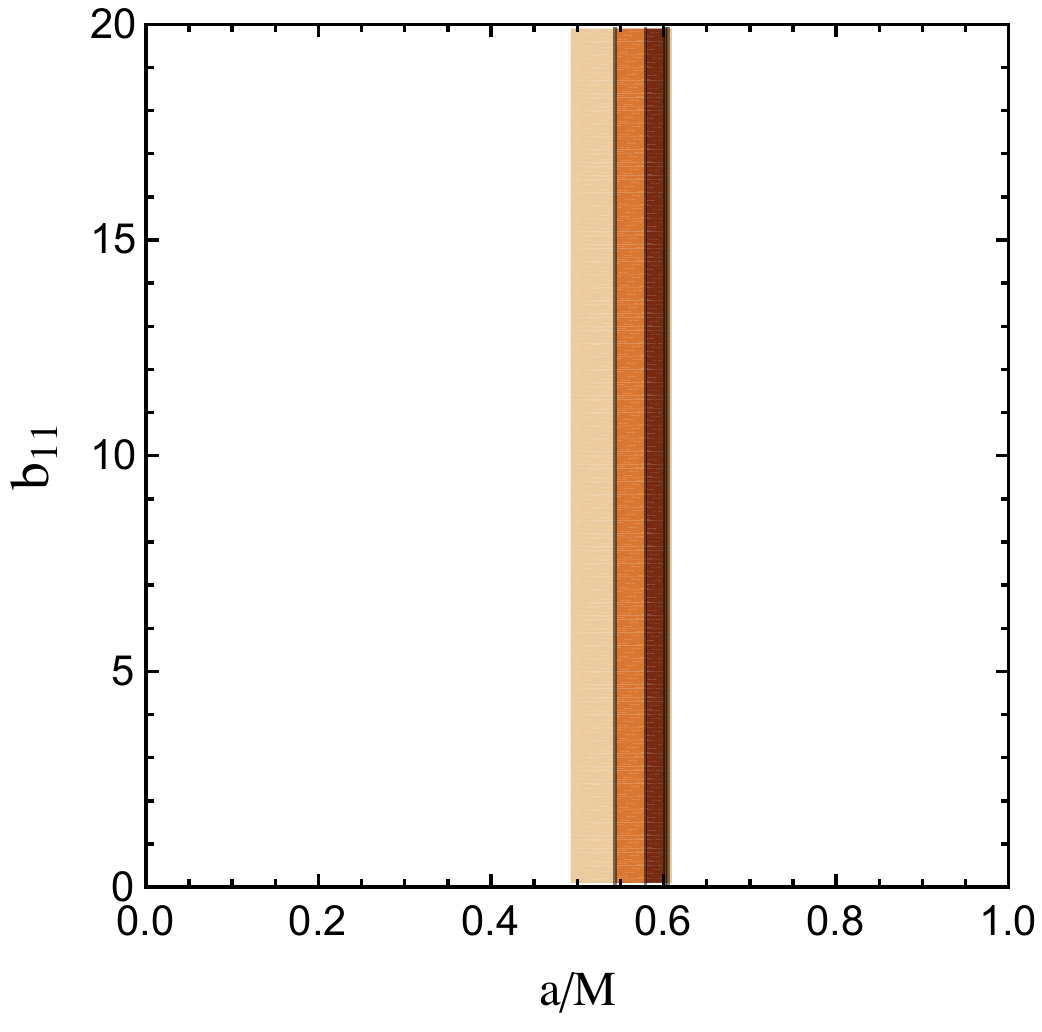}
\end{center}
\vspace{-0.3cm}
\caption{Contour map of thermal spectra of accretion disks. Contour map of parameter $b_8$ (top left panel), $b_9$ (top right panel), $b_{10}$ (bottom left panel), $b_{11}$ (bottom right panel). The reference model is a Kerr BH with spin 0.6. The other parameter values are $M = 10 M_{\odot}$,  $\dot M = 2 \times 10^{18}$ g/s, $D = 10$ kpc, and $i = 45^{\circ}$. See the text for more details. \label{con2}}
\end{figure*}

\begin{figure*}
\vspace{0.4cm}
\begin{center}
\includegraphics[type=pdf,ext=.pdf,read=.pdf,width=8.0cm]{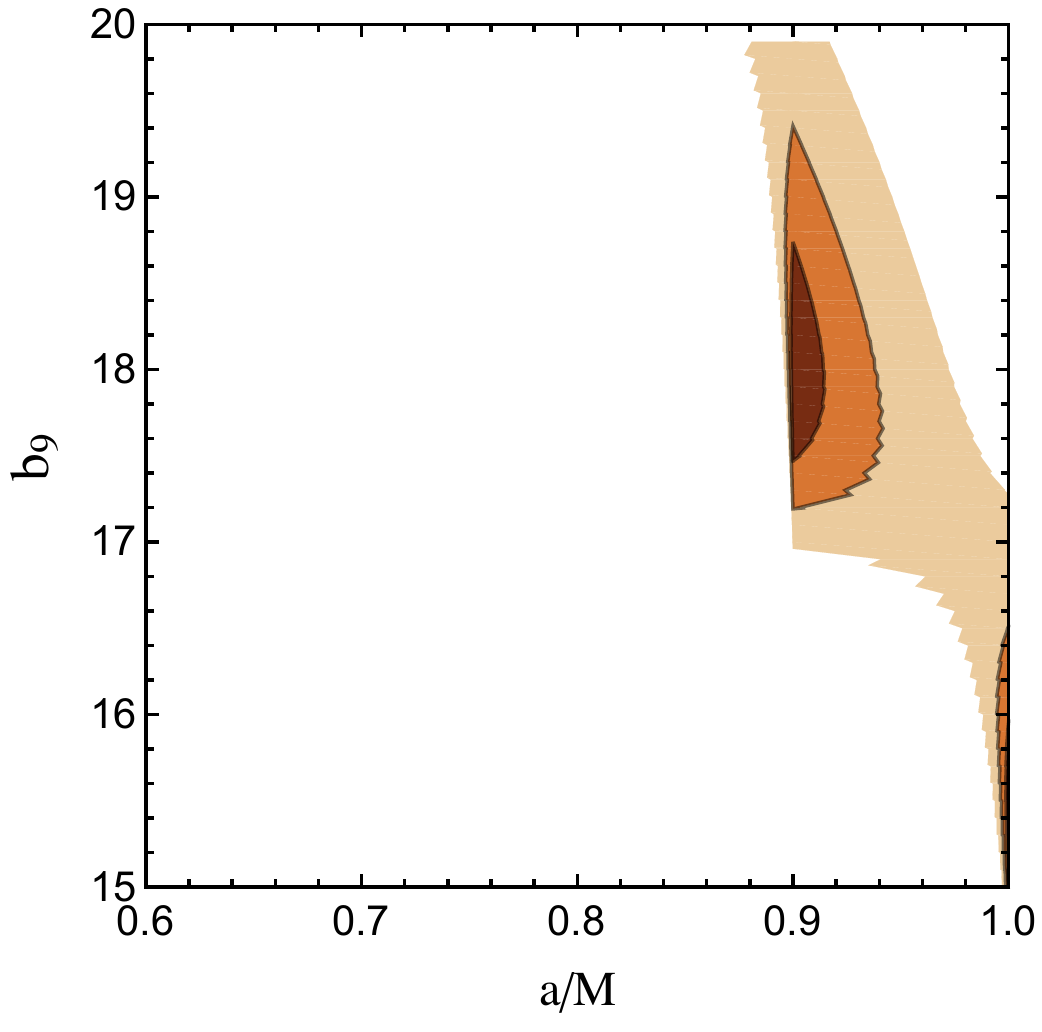}
\end{center}
\vspace{-0.3cm}
\caption{Contour map of thermal spectra for parameter $b_9$. The reference model is a non-Kerr BH with $M = 10 M_{\odot}$,  $\dot M = 2 \times 10^{18}$ g/s, $D = 10$ kpc, $i = 45^{\circ}$, $a_* = 0.9,$ and $b_9 = 18$. In these simulations I have assumed $f_{col} = \Upsilon = 1$. See text for more details.\label{b9gh}}
\end{figure*}

The results of my simulation of thermal spectra are shown in Figs.\ref{spectra}-\ref{b9gh}. In Fig.~\ref{spectra} I have drawn the impact of deformation parameter $b_i$ on the thermal spectra of the accretion disk for different values of the model parameters. The values of the model parameters are $M = 10 M_{\odot}$, $\dot M = 2 \times 10^{18}$ g/s, $D = 10$ kpc, $i = 45^{\circ}$, $a_* = a/M = 0.9,$. In each curve one $b_i$ is equal to 15 and all others are equal to 1. In these simulations I have assumed $f_{col} = \Upsilon = 1$.
As we see in the plot, the parameter $b_7$ makes the spectra harder than the Kerr one on increasing. By increasing the parameter $b_4, b_9$ and $b_{10}$ the spectra get softer than Kerr case. Parameters $b_2, b_5, b_8,$ and $b_{11}$ are nearly similar to Kerr one. The overall structure of the plots does not change for different sets of parameters.
The impact of all other parameters is studied in \cite{nk1,con1,con3}.

Now, in order to compare my results, I use  the$\chi^2$ approach. I compare the spectra produced in Kerr spacetime with one expected from the GB background. The $\chi^2$ is defined as follows:
\be
\chi^2 (a_*, b_j)= \frac{1}{n} \sum_{i=1}^n \frac{N_i^{GB} - N_i^{Kerr}}{\sigma_i^2}
\ee
where summation is performed over $n$ sampling energies $E_i$. $N_i^{Kerr}$ and $N_i^{GB}$ are the photon fluxes in the energy bin, $[E_i, E_i + \Delta E]$, for Kerr and GB metric. The error, $\sigma_i$, is considered to be $15\%$ of the photon flux $N_i^{Kerr}$. For these results, $M = 10 M_{\odot}$,  $\dot M = 2 \times 10^{18}$ g/s, $D = 10$ kpc, $i = 45^{\circ}$ and $a_* = 0.6$. The contours are presented in Figs.~\ref{con1} and~\ref{con2}. 
 The brown, orange and lighter orange are for $1 \sigma, 2 \sigma$ and $3 \sigma$ (68\%, 95\% and 99.7\%) confidence level, respectively.
 If the contour plots would be closed, this means we can constrain the parameters and there is no degeneracy, but, here, the contours are not closed in Figs.~\ref{con1} and~\ref{con2}. So we cannot constrain the parameters. In other words, there is degeneracy between measurements of the parameters $b_i$ and measurement of the spin parameter, so that the parameters $b_i$ cannot be constrained. Degeneracy means one cannot distinguish the properties of these spectra from spectra of a disk around a Kerr BH.

This degeneracy is a problem as regards measuring the spin of astrophysical BHs. So far, spin measurement is based on the Kerr BH hypothesis but as we see from the result of this study with the continuum-fitting method, the same electromagnetic observation (thermal spectra here) can be reproduced by non-Kerr spacetime as an example of GB spacetime with different spin and deformation parameters. 

I also consider the case for higher values of the spin, such as $0.9$, as reference model but still the results are degenerate. Then, I consider the reference BH not to be a Kerr BH. I consider a GB BH as reference BH in a contour study with spin $0.9$ and $b_i = 18$. In this case, only the parameter $b_9$ may be constrained because the contour looks closed, as we see in Fig.~\ref{b9gh}.
The parameter $b_{10}$ is harder to constrain is this case as well.

As the result of other tests with electromagnetic radiation, considering triplet frequency observations of GRO J1655-40, we find that the GB parameter $b_2$ is $2.2_{-0.523}^{+0.395}$ but the other parameters cannot be constrained~\cite{qpoma}. If we assume we have 10 observations in the future, we can constrain more parameters, we find that $b_2, b_9, b_{10}$ and $b_{11}$ can be well constrained, while $b_4, b_5, b_7$ and $b_8$ cannot.
With X-ray reflection spectroscopy, known as the iron line method, 200 ks observations with future observational facilities such a LAD-eXTP can constrain all parameters except for $b_{11}$.  $b_{11}$ leaves very weak impact on the iron line. A BH shadow method shows a weak impact on the parameters $b_2, b_8, b_9$ and $b_{10}$. There is no impact on the shadow boundary by the parameters $b_4, b_5, b_7$ and $b_{11}$. An iron line reverberation mapping study of GB spacetime can constrain all parameters except $b_{11}$. $b_4$ also is harder to constrain~\cite{qpoma}. \\
As we see the BH shadow can constrain some parameters but can not constrain others. Iron line studies either with time-integrated analysis or with reverberation can constrain all with exception of $b_{11}$. But a mock observation of 10 BHs can constrain $b_{11}$ in addition to the other parameters.\\
Unfortunately, due to degeneracy of the parameters in the continuum-fitting method, we cannot further restrict GB parameters using this method. This might be a general problem as regards measuring the spin of the BH.

A combination of different observational methods may help to constrain the deformation parameters and break the degeneracy of the parameters.

%%%%%%%%%%%%%%%%%%%%%%%%%%%%%%%

\section{Summary and conclusions \label{summary}}

The continuum-fitting method is a method to study thermal spectra. The disk should be geometrically thin and optically thick. Also this study is valid for stellar-mass black holes. 

Furthermore, parametrization of the Kerr BH hypothesis is one way to study deviations from the predictions of GR and provide constraints on the deviation from GR. 

In this work I use the continuum-fitting method to check whether it can constrain the GB metric parameters introduced in~\cite{gbma}. The GB metric contains 11 deformation parameters, which the Kerr case recovers when all deformation parameters are set to one. I first compute the thermal spectra for different $b_i$ parameters. My results show that by increasing parameter $b_7$ the spectra get harder. Also by increasing the parameters $b_4, b_9, b_{10}$ the spectra get softer and impact of the parameters $b_2, b_5, b_8,$ and $b_{11}$ is very close to the Kerr one. 

I then consider my simulation as observational data and try to study possible constraints of the deformation parameters of GB spacetime using the $\chi^2$ approach. My contour studies show that all deformation parameters are degenerate. However, for the case of a high spin value with GB BH as reference, the parameter $b_9$ may be constrained. 
This degeneracy means the observation with specific spin and the deformation parameter can reproduce the spectra of the Kerr case. This is a problem as regards of measuring the spin of astrophysical black holes and constrain possible deviations from GR. Also as a result of this degeneracy, in order to verify the Kerr hypothesis, any deviation from the Kerr case should be vanish.

Different observational tests such as the BH shadow method, X-ray reflection spectroscopy, iron line reverberation mapping and QPO introduce different impacts of the parameters. Combination of different methods may help to break the degeneracies.

%%%%%%%%%%%%%%%%%%%%%%%%%%%%%%%

{\bf Acknowledgments}

This work is supported by School of Astronomy, Institute for Research in Fundamental Sciences (IPM), Tehran, Iran.

%%%%%%%%%%%%%%%%%%%%%%%%%%%%%%%


\begin{thebibliography}{}
\bibitem{qpoma} 
  The more general and detailed calculations will be presented in a forthcoming publication by the author.
\end{thebibliography}
\end{document}